\documentstyle[11pt]{article}
\input epsf
\begin{document}
\thispagestyle{empty}
\begin{center}
\null\vspace{-1cm} {\footnotesize {\tt }}\hfill VACBT/05/13,
GNPHE/05/13 and IC/2005\\ \vspace{1cm}
\medskip
\vspace{2.5 cm} {\large \textbf{The Moyal Momentum Algebra}}\\
\vspace{2 cm} \textbf{A. Boulahoual and {M.B.
SEDRA}\footnote{sedra@ictp.it}}
\\
{\small \ International Centre for Theoretical Physics, Trieste,
Italy.}\\ {\small \ Virtual African Center For Basic Sciences and
Technology, VACBT,\\ Focal point: Lab/UFR-Physique des Haute
Energies, Facult´e des Sciences, Rabat, Morocc}.\\ {\small \
Groupement National de Physique de Hautes Energies, GNPHE, Rabat,
Morocco,}\\{\small \ \ \ } {\small \ \ \ }{\small \ Universit\'{e}
Ibn Tofail, Facult\'{e} des Sciences, D\'{e}partement de
Physique,}\\ {\small \ \ \ }{\small \ Laboratoire de Physique de
la Mati\`{e}re et Rayonnement (LPMR), K\'{e}nitra, Morocco}\\
\end{center}
\vspace{0.5cm} \centerline{\bf Abstract} \baselineskip=18pt
\bigskip
{\small We introduce in this short note some aspects of the Moyal
momentum algebra introduced first by Das and Popowicz and that is
denoted by Mm algebra. Our interest on this algebra is motivated
by the central role that it can play in the formulation of
integrable models and in higher conformal spin theories. }

\hoffset=-1cm \textwidth=11,5cm \vspace*{1cm}
\hoffset=-1cm\textwidth=11,5cm \vspace*{0.5cm}

\newpage

\section{\protect\small Introduction}

{\small In the past few years, there has been a growth in the interest in
non-commutative geometry (NCG), which appears in string theory in several
ways [1]. Much attention has been paid also to field theories on NC spaces
and more specifically Moyal deformed space-time, because of the appearance
of such theories as certain limits of string, D-brane and M-theory [2]. One
of the strong points of the NCG framework is its richness and also the fact
that we can recover all the well known standard results just by requiring
the vanishing of the deformed parameter which means also the vanishing of
noncomutativity. Note that the passage from commutative to noncommutative
space time is achieved by replacing the ordinary commutative product, in the
space of smooth functions on $R^2$ with coordinates $x, t$, by the
noncommutative associative $\star$- product [3]. Moyal deformation is
applied also to Lax equations and supersymmetric KdV hierarchies [4,5].%
\newline
The aim of this letter is to introduce some aspects of the Moyal
momentum algebra that we call the Moyal Momentum (Mm)algebra [4]
shown to play, in the NCG framwork, an important role in
formulating integrable models and higher conformal spin theories
in a systematic way [6]. }

\section{\protect\small The Moyal Momentum algebra}

{\small Using our convention notations [7], we denote this algebra by ${%
\widehat \Sigma} (\theta)$. This is a non-commutative space based on
arbitrary momentum Lax operators and which decomposes as:
\begin{equation}
{\widehat \Sigma} (\theta) = \oplus _{r\leq s} \oplus _{m \in Z}{\widehat
\Sigma}_{m}^{(r,s)},
\end{equation}
where ${\widehat \Sigma}_{m}^{(r,s)} (\theta)$ is the space of $\theta$-Lax
operators of fixed conformal spin m and degrees (r,s) given by
\begin{equation}
{\cal L}_{m}^{(r,s)}(u)=\sum _{i=r}^{s}u_{m-i}(x)\star p^i.
\end{equation}
These are $\theta$-differentials whose operator character is inherited from
the star product law defined as follows: }

{\small
\begin{equation}
f(x,p)\star g(x,p)= \sum_{s=0}^{\infty} \sum_{i=0}^{s}{\frac{\theta ^s}{s!}}
(-)^{i} c _{s}^{i}
(\partial_{x}^{i}\partial_{p}^{s-i}f)(\partial_{x}^{s-i}\partial_{p}^{i}g),
\end{equation}
with $c _{s}^{i}=\frac {s!}{i!(s-i)!}$ and $f(x,p)$ are arbitrary functions
on the two-dimensional phase space.\newline
Now, it is important to precise how the momentum operators act on arbitrary
functions $f(x,p)$ via the star product. Performing computations based on
the relation (3), we find the following $\theta$- Leibniz rules: }

{\small
\begin{equation}
p^{n} \star f(x,p) = \sum _{s=0}^{n} \theta ^{s} c_{n}^{s} f^{(s)}(x,p)
p^{n-s},
\end{equation}
and
\begin{equation}
p^{-n} \star f(x,p) = \sum _{s=0}^{\infty} (-)^{s} \theta ^{s} c_{n+s-1}^{s}
f^{(s)}(x,p) p^{-n-s},
\end{equation}
}

{\small with $n\in N$ and $f^{(s)}=(\partial_{x}^{s}f)$ is the prime
derivative.\newline
}

{\small We find also that the Moyal bracket defined as [8]: }

{\small
\begin{equation}
\{f(x,p), g(x,p)\}_ {\theta} =\frac {f \star g - g \star f}{2\theta}
\end{equation}
}

{\small is subject to the following expressions }

{\small
\begin{equation}
\begin{array}{lcl}
\{p^n, f\}_{\theta} & = & \sum _{s=0}^{n} \theta ^{s-1} c_{n}^{s}\{\frac
{1-(-)^s}{2}\} f^{s} p^{n-s}, \\
\{p^{-n}, f\}_{\theta} & = & \sum _{s=0}^{\infty} \theta ^{s-1}
c_{s+n-1}^{s}\{\frac {(-)^{s}-1}{2}\} f^{s} p^{-n-s}.
\end{array}
\end{equation}
}

{\small Having defined the Moyal Momentum  algebra, we present
here bellow, some remarkable properties.\newline {\bf P1}: We give
here bellow the conformal dimensions of objects used in this
study\newline $[x]=-1$, $[p]=[\partial_x]=1$, $[\theta]=0$,
$[u_m]=m$, $[\widehat Res]=[Res]=1$\newline {\bf P2}: The momentum
operators $p^i$ satisfy
\begin{equation}
p^n \star p^m =p^{n+m}.
\end{equation}
ensuring the suspected rule
\begin{equation}
p^n \star (p^{-n}\star f(x,p)) = f(x,p).
\end{equation}
\newline
{\bf P3}: We should note that formulas (4-5,7) are computed for integers
values $n =0, 1, 2, 3,...$. Now, what happens if half integers $n={\pm \frac
{1}{2}}$,${\pm\frac {3}{2}}$, ${\pm\frac {5}{2}},...$ or arbitrary
fractional powers were allowed?\newline
Using the general formula (3) and performing algebraic computations we find
the following formulas:
\begin{equation}
{p^{\frac{a}{2}} \star f(x,p)} = {\sum _{s=0}^{\infty}}{\Pi _{j=0}^{s-1}}({%
\frac{a}{2}}-j) {\frac {\theta ^{s}}{s!}}f^{(s)}(x,p) p^{\frac{a}{2}-s},
\end{equation}
describing the $\star$-product action of half integer powers of the momentum
operators on the phase space with
\begin{equation}
{\Pi _{j=0}^{s-1}}({\frac{a}{2}}-j)=(\frac{a}{2})(\frac{a}{2}-1)(\frac{a}{2}%
-2)...(\frac{a}{2}-s+1).
\end{equation}
Moreover, from eq.(10) one can recover eqs.(4-5) once even values $a=2n$,
(for n positive or negative) are considered. We can then conclude that these
equations as well as eq.(8) are special cases of (10). The more general
situation consist in considering fractional objects type $p^{\frac{a}{b}}$,
we have
\begin{equation}
{p^{\frac{a}{b}} \star f(x,p)} = {\sum _{s=0}^{\infty}}{\Pi _{j=0}^{s-1}}({%
\frac{a}{b}}-j) {\frac {\theta ^{s}}{s!}}f^{(s)}(x,p) p^ {\frac{a}{b}-s},
\end{equation}
}

{\small Few remarks are in order. If eqs.(4-5) are associated to
`bosonic' behavior of the Moyal Momentum  algebra $\widehat
\Sigma$, the half conformal spin object ${p^{\frac{1}{2}}\star
u_{0}} \in {\widehat \Sigma^{({\frac {1}{2}}, {\frac
{1}{2}})}_{\frac{1}{2}}}$, for example, may leads us to interpret
naively this object as being of fermionic character. The same
judgment is made for the fractional object ${p^{\frac{a}{b}}\star
u_{0}} \in {\widehat \Sigma ^{(\frac {a}{b},
\frac{a}{b})}_{\frac{a}{b}}}$ having fractional conformal
spin.\newline The obviousness of this correspondence with
statistics is only impression. In fact, for example for objects
like $p^{\frac{1}{2}}\star u_{0}$, even though they present half
conformal spins, they are not fermionic objects because
$p^{\frac{1}{2}} \star p^{\frac{1}{2}}=p$ and a fermionic object
should be nilpotent. In order to attribute a fermionic behavior to
our momentum operators, one have to consider a graded phase space
parametrized by the set of variables ($x$, $p$, $\eta$ and
$p_{\eta}$) where $\eta$ is the Grassmann variable and $p_{\eta}$
the corresponding nilpotent conjugate momenta given by
$p_{\eta}^2=0$ [5].\newline Nevertheless, it would be very
interesting to look for the contribution and the meaning of these
fractional powers of momenta in the framework of KdV hierarchy
equations and also in the building of the GD second Hamiltonian
structure.\newline {\bf P4}: The space ${\widehat \Sigma}(\theta)$
may decomposes into the underlying sub-algebras as
\begin{equation}
\begin{array}{lcl}
{\widehat \Sigma} (\theta) & = & \oplus _{r\leq s} {\widehat \Sigma}^{(r,s)}
\\
& = & \oplus _{r\leq s} \oplus _{m \in Z}{\widehat \Sigma}%
_{m}^{(r,s)}(\theta) \\
& = & \oplus _{r\leq s} \oplus _{m \in Z} \oplus _{k=r}^{s} {\widehat \Sigma}%
_{m}^{(k,k)}(\theta)
\end{array}
\end{equation}
where ${\widehat \Sigma}_{m}^{(k,k)}(\theta)$ is generated by elements type $%
u_{m-k}\star p^k$ or $p^k \star v_{m-k} $. \newline
{\bf P5}: Using the $\theta$-Leibniz rule, we can write, for fixed value of
k:
\begin{equation}
{\widehat \Sigma}_{m}^{(k,k)} \equiv \Sigma _{m}^{(k,k)} \oplus \theta
\Sigma _{m}^{(k-1,k-1)} \oplus \theta^{2}{\Sigma}_{m}^{(k-2,k-2)} \oplus...
\end{equation}
where $\Sigma _{m}^{(k,k)}$ is the standard one dimensional sub-space,
containing the prototype objects $u_{m-k} p^k$ that we consider as the $%
(\theta=0)$-limit of ${\widehat \Sigma}_{m}^{(k,k)}$. \newline
{\bf P6}: The space ${\widehat \Sigma}_{m}^{(0,0)} \equiv \Sigma
_{m}^{(0,0)} $ is nothing but the ring of analytic fields $u_{m}$ of
conformal spin ${m \in Z}$. With respect to this definition, the subspace ${%
\widehat \Sigma}_{m}^{(k,k)}$ can be written formally as:
\begin{equation}
{\widehat \Sigma}_{m}^{(k,k)} \equiv p^{k} \star \Sigma _{m-k}^{(0,0)}.
\end{equation}
{\bf P7}: We can easily check that, in general, the space ${\widehat \Sigma}%
_{m}^{(k,k)}$ is not closed under the action of Moyal bracket (6) since we
have:
\begin{equation}
\{.,.\}_\theta : {\widehat \Sigma}_{m}^{(r,s)} \star {\widehat \Sigma}%
_{m}^{(r,s)} \rightarrow{\widehat \Sigma}_{2m}^{(r,2s-1)}
\end{equation}
Imposing the closure, one gets strong constraints on the integers m, r and s
namely
\begin{equation}
m=0 \\
, r\leq s \leq 1
\end{equation}
Under these constraint equations, the sub-spaces ${\widehat \Sigma}%
_{m}^{(r,s)}$ exhibit then a Lie algebra structure since the $\star$-product
is associative. \newline
{\bf P8}: The sub-space ${\widehat \Sigma}_{m}^{(r,s)}$ is characterized by
the existence of a residue operation that we denote as ${\widehat Res}$ and
which acts as follows
\begin{equation}
\begin{array}{lcl}
\widehat{Res}(u_{k}\star p^{-k}) & = & (u_{k} \star p^{-k}) \delta _{k-1,0}
\\
& = & u_{1}\delta _{k-1,0}
\end{array}
\end{equation}
This result coincides with the standard residue: $Res$, acting on the
sub-space ${\Sigma}_{m}^{(r,s)}$:
\begin{equation}
Res (u_1 . p^{-1})=u_1
\end{equation}
We thus have two type of residues $\widehat Res$ and $Res$ acting on two
different spaces ${\widehat \Sigma}_{m}^{(r,s)}$ and ${\Sigma}_{m}^{(r,s)}$
but with value on the same ring ${\Sigma}_{m+1}^{(0,0)}$. This Property is
summarized as follows:
\begin{equation}
\begin{array}{lcl}
{\widehat \Sigma}_{m}^{(r,s)} \longrightarrow _{\theta =0} {\Sigma
_{m}^{(r,s)}} &  &  \\
\hspace{0,2cm}{_{\widehat Res}} {\large \searrow} \hspace{1cm} {\large %
\swarrow}_{{Res}} &  &  \\
\hspace{1,15cm} \Sigma _{m+1}^{(0,0)} &  &
\end{array}
\end{equation}
}

{\small We learn from this diagram that the residue operation exhibits a
conformal spin quantum number equal to 1.\newline
{\bf P9}: With respect to the previous residue operation, we define on ${%
\widehat \Sigma}$ the following degrees pairing
\begin{equation}
(.,.): {\widehat \Sigma}_{m}^{(r,s)} \star {\widehat \Sigma}%
_{n}^{(-s-1,-r-1)} \rightarrow {\Sigma}_{m+n+1}^{(0,0)}
\end{equation}
such that
\begin{equation}
({}_{m}^{(r,s)}(u),\tilde {}_{n}^{(\alpha ,\beta)}(v))=\delta_{\alpha
+s+1,0}\delta_{\beta+r+1,0}Res({}_{m}^{(r,s)}(u)\star \tilde {}_{n}^{(\alpha
,\beta)}(v)),
\end{equation}
showing that the spaces ${\widehat \Sigma}_{m}^{(r,s)}$ and ${\widehat
\Sigma }_{n}^{(-s-1 ,-r-1)}$ are $\widehat Res$-dual as ${\Sigma}%
_{m}^{(r,s)} $ and ${\Sigma }_{n}^{(-s-1 ,-r-1)}$ are dual with respect to
the $Res$-operation.\newline
Other important properties with possible applications will be considered in
a forthcoming work. \newline
}

{\small {\bf References} }

\begin{enumerate}
\item[{[1]}]  {\small A. Connes, Noncommutative geometry, Academic Press
(1994),\newline
A. Connes, M.R. Douglas, A. Schwarz, Noncommutative geometry and matrix
theory: Compactification on tori, JHEP 02(1998) 003, [hep-th/9711162] and
references there in. }

\item[{[2]}]  {\small N. Seiberg and E. Witten, String theory and
non-commutative geometry, JHEP 09(1999) 032. }

\item[{[3]}]  {\small M. Kontsevitch, Deformation quantization of Poisson
manifolds I, [q-alg/9709040]\newline
D.B. Fairlie, Moyal Brackets, Star Products and the Generalized Wigner
Function, [hep-th/9806198]\newline
D.B. Fairlie, Moyal Brackets in M-Theory, Mod.Phys.Lett. A13 (1998) 263-274,
[hep-th/9707190],\newline
C. Zachos, A Survey of Star Product Geometry, [hep-th/0008010],\newline
C. Zachos, Geometrical Evaluation of Star Products, J.Math.Phys. 41 (2000)
5129-5134, [hep-th/9912238],\newline
C. Zachos, T. Curtright, Phase-space Quantization of Field Theory,
Prog.Theor. Phys. Suppl. 135 (1999) 244-258, [hep-th/9903254]. }

\item[{[4]}]  {\small A. Das and Z. Popowicz, Properties of Moyal-Lax
Representation, Phys.Lett. B510 (2001) 264-270, [hep-th/0103063. }

\item[{[5]}]  {\small A. Das and Z. Popowicz, Supersymmetric Moyal-Lax
Representations, J. Phys. A 34(2001) 6105 and hep-th/0104191. }

\item[{[6]}]  {\small Ming-Hsien Tu, $W_{n}^{\theta }$ algebra associated
with the Moyal KdV hierarchy, Phys.Lett. B508 (2001) 173-183, \newline
Ming-Hsien Tu, Niann-Chern Lee, Yu-Tung Chen, Conformal Covariantization of
Moyal-Lax Operators, J.Phys. A35 (2002) 4375.\newline
}

\item[{[7]}]  {\small E.H.Saidi and M.B.Sedra, On the Gelfand Dickey algebra
$GD(sl_{n})$ and the $W_{n}$-symmetry, J.Math.Phys.35(6), June 1994,\newline
M.B.Sedra, On The huge Lie superalgebra of pseudo superdifferential
operators and super KP Hierarchies, J.Math.Phys.37(1996)3483.\newline
}

\item[{[8]}]  {\small H. Groenewold, Physica 12(1946)405,\newline
J.E. Moyal, Proc. Cambridge Phil. Soc. 45(1949)90. }
\end{enumerate}

\end{document}